# Sulfonylamide-Based Ionic Liquids for High-Voltage Potassium-Ion Batteries with Honeycomb Layered Cathode Oxides

*Kazuki Yoshii[a], Titus Masese[a,b], Minami Kato[a], Keigo Kubota[b], Hiroshi Senoh[a] and Masahiro Shikano[a]*

[a] Research Institute of Electrochemical Energy, National Institute of Advanced Industrial Science and Technology (AIST), 1–8–31 Midorigaoka, Ikeda, Osaka 563–8577, JAPAN
[b] AIST–Kyoto University Chemical Energy Materials Open Innovation Laboratory (ChEM–OIL), Sakyo–ku, Kyoto 606–8501, JAPAN

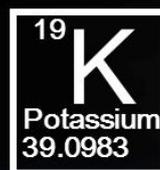

## Abstract:

The world is at the cusp of a new era where pivotal importance is being attached to the development of sustainable and high-performance energy storage systems. Potassium-ion batteries are deemed not only as cheap battery candidates, but also as the penultimate high-voltage energy storage systems within the monovalent-cation chemistries. However, their performance and sustainability are undermined by the lack of suitable electrolytes for high-voltage operation particularly due to the limited availability of cathode materials. Here, the potential of ionic liquids based on potassium bis(trifluoromethanesulfonyl)amide (KTFSA) as high-voltage electrolytes is presented by assessing their physicochemical properties, along with the electrochemical properties upon coupling with new high-voltage layered cathode materials. These ionic liquids demonstrate a lower redox potential for potassium dissolution / deposition (with a wide voltage tolerance of around 6.0 V), placing them as feasible and safe electrolytes for high-voltage potassium-ion battery configuration. This is proven by matching this electrolyte with new high-voltage layered cathode compositions, demonstrating stable electrochemical performance. The present findings of electrochemically stable ionic liquids based on potassium bis(trifluoromethanesulfonyl)amide will bolster further advancement of high-performance cathode materials, whose performance at high-voltage regimes were apparently restricted by the paucity of suitable and compatible electrolytes.

# 1. Introduction

An exigent need exists to meet the world's gigantic energy demands in a sustainable manner with low environmental impact. Lithium-ion battery technology, previously seen as a universal choice to cater for the escalating energy demands, is facing uncertainty due to dwindling natural reserves.[1] Thus, the development of disruptive low-cost battery technologies meant for largescale and capacious energy storage systems is utterly desirable. In this context, batteries relying on the abundant potassium element, as charge-carriers, are deemed not only as lucrative alternatives, but also as the penultimate high-voltage energy storage contender systems.[2-4] Aside from incurring a significant reduction in potassium-ion (K-ion) battery design (through the use of graphite as a standard anode material and aluminum foils as current collectors for the anode as well), potassium marks a redox potential close to or even lower than lithium in non-aqueous electrolytes.[5] This suggests that the emergence of a high-voltage K-ion battery system that is cheap, could be anticipated.

The development of a high-voltage K-ion battery system is, however curbed by the scarcity of cathode materials that can accommodate the large K-ions, along with suitable electrolytes that can operate safely and stably at high-voltage regimes. Although several high-voltage cathode materials have been reported, which is in austere contrast to the considerable progress made for anode materials,[6] a challenge still lurks to find compatible and stable electrolytes particularly for high-voltage cathode materials. **Figure 1a** shows a majority of the reported cathode materials along with the attained average voltages. It is evident that few high-voltage cathodes exist (exhibiting average working voltages of close to or beyond 4 V); Prussian organic moieties and polyanionic compounds such as $KVOPO_4$, $KVP_2O_7$, $KVPO_4F$,[6-8] amongst others dominating as high-voltage cathode candidates. Potassium-based oxides tend to show low average voltage; however, this prevailing notion was recently obliterated by the design of tellurium-doped $K_{2/3}Ni_{2/3}Te_{1/3}O_2$ (or equivalently as $K_2Ni_2TeO_6$, for simplicity) and $K_{2/3}Ni_{1/3}Co_{1/3}Te_{1/3}O_2$ ($K_2NiCoTeO_6$) as high-voltage layered cathode materials (Figure 1a).[6, 7, 9-13] Despite the alluring prospects of using these high-voltage cathode materials to develop a high-voltage battery system, instability of organic electrolytes at high-voltage operation coupled with the high reactivity of potassium metal anode continues to cast doubt on the safety of electrolytes based on organic solvents.[14]

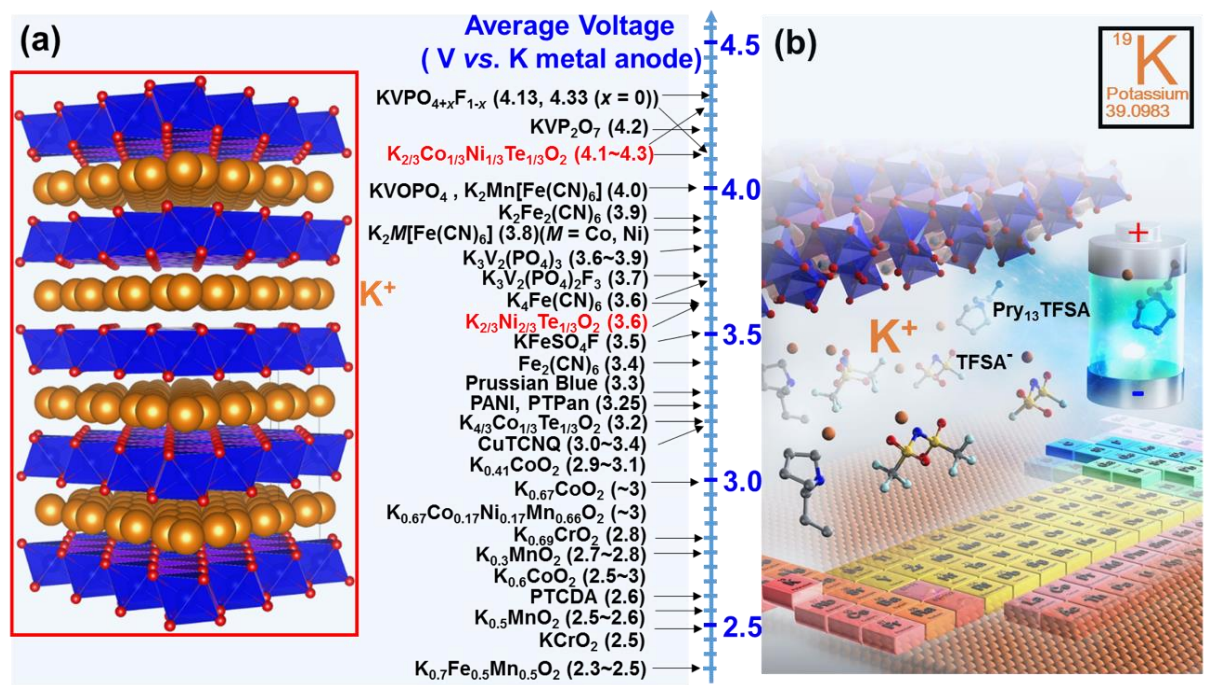

**Figure 1.** Schematic illustration of the average voltages attained by cathode materials that can be coupled to stable ionic liquids based on bis(trifluoromethanesulfonyl)amide for the design of high-energy-density potassium-ion battery. **(a)** Reported cathode materials highlighting the honeycomb layered high-voltage tellurate compounds (in red), solid-solution derivatives of which we focus on, along with their typical honeycomb crystal framework entailing potassium atoms sand-witched between transition metal layers. **(b)** High-voltage battery configurations that can be envisioned by coupling high voltage honeycomb-layered frameworks with suitable anodes and ionic liquid based on potassium bis(trifluoromethanesulfonyl)amide (KTFSA) salt in 1-methyl-1-propylpyrrolidinium bis(trifluoromethanesulfonyl)amide ($Pyr_{13}TFSA$) ionic liquid.

Ionic liquids are a burgeoning class of safe electrolytes that have particularly been sought to address the stability of cathode materials under high-voltage operation conditions in batteries such as the lithium-, sodium-, magnesium- and aluminum-ion technologies,[15-18] but not yet as prolific in the nascent potassium-ion battery research. Ionic liquids, which comprise (in)organic anions and organic cations, are characterized by a suite of unique properties; prime amongst them being their good thermal stability, low flammability and low volatility. These virtues guarantee improved safety with the use of ionic liquids over organic solvent electrolytes. Moreover, the inherently wide electrochemical window endows ionic liquids with great prospects in high-voltage battery operation.

Studies relating to ionic liquids that contain potassium salts are, unfortunately, still at their infancy. Yamamoto and co-workers reported ionic liquids based on potassium bis(fluorosulfonyl)amide (hereafter denoted as KFSA) in 1-methyl-1- propylpyrrolidinium bis(trifluoromethanesulfonyl)amide (hereafter $Pyr_{13}FSA$) as potential safe electrolytes for high-voltage potassium-ion battery, demonstrating a remarkably wide electrochemical window (> 5.72 V) in conjunction with reasonable ionic conductivity.[19] Beltrop and co-workers further elegantly demonstrated ionic liquids based on potassium bis(trifluoromethanesulfonyl)amide (KTFSA) in 1-butyl-1-methylpyrrolidinium bis(trifluoromethanesulfonyl)amide ($Pyr_{14}TFSA$) as a feasible electrolyte for stable performance of potassium dual-ion battery.[20] Recently, we found that ionic liquids based on KTFSA salt in 1-propyl-1-methylpyrrolidinium TFSA ($Pyr_{13}TFSA$) demonstrate stable performance when coupled to high-voltage layered oxides such as $K_{2/3}Ni_{2/3}Te_{1/3}O_2$.[9, 10] However, the physicochemical properties of ionic liquids based on KTFSA salt remain to be expounded; a necessary prelude to assessing their potential role as safe electrolytes for high-voltage cathode materials we envision in **Figure 1b**.

Described herein are the physicochemical and electrochemical properties of ionic liquids based on KTFSA salt in $Pyr_{13}TFSA$, to address the fundamental aspects about the functionalities of related ionic liquids that we identify. It is revealed that the redox potential for potassium dissolution / deposition using KTFSA ionic liquids is lower than for their lithium and sodium counterparts, resulting in a wider electrochemical window (around 6.0 V). In addition, stable performance of these ionic liquids is shown by coupling to new high-voltage layered cathode compositions that we synthesize. Altogether this work brings to the limelight, ionic liquids based on KTFSA salt as potent and safe electrolytes for the design of high-voltage potassium-ion battery.

## 2. Results and Discussion

### 2.1. Physicochemical and Electrochemical Properties of Potassium Bis(Trifluoromethanesulfonyl)amide (KTFSA)-based Ionic liquids

Potassium bis(trifluoromethanesulfonyl)amide (KTFSA) and two types of TFSA-based ionic liquids (namely, 1-propyl-1-methylpyrrolidinium bis(trifluoromethanesulfonyl)amide ($Pyr_{13}TFSA$) and 1-ethyl-3-methylimidazolium bis(trifluoromethanesulfonyl)amide (EMITFSA)), were mixed in an argon-purged glovebox at room temperature. Further details are provided in the **Experimental** section. While the lithium and sodium analogues (*i.e.*, LiTFSA and NaTFSA) can dissolve to concentrations close to or exceeding 1.0 M (mol dm$^{-3}$) in these ionic liquids at room temperature, KTFSA dissolves only to a concentration of 0.6 M in $Pyr_{13}TFSA$ and 0.7 M in EMITFSA, respectively. The lower solubility of KTFSA is ascribed to the large ionic size of potassium cations ($K^+$) and its inherently lower Lewis acidity compared with Li and Na ions.

To assess the phase stability of KTFSA-based ionic liquids, differential scanning calorimetry (DSC) measurements were conducted (as shown in **Figures S1a and S1c**). An endothermic peak manifesting the melting point ($T_m$) could be clearly seen in $Pyr_{13}TFSA$. $T_m$ shifts to lower temperatures with the addition of KTFSA, and another peak emerges, suggesting that the interaction between the $K^+$ and $TFSA^-$ from $Pyr_{13}TFSA$ alters the crystal structure of the ionic liquid to a metastable state.[21] A comparison of the phase transition of $M$TFSA ($M$ = Li, Na, and K)-based ionic liquids adopting the same concentration (see **Figures S1b and S1d**), reveals that the melting point ($T_m$) decreases in the sequence: KTFSA > LiTFSA > NaTFSA. This trend presumably is influenced by the order of the melting temperatures of $M$TFSA salts.[22] No marked difference in the trend of phase transition between $Pyr_{13}$TFSA and EMITFSA-based system was observed. We are indeed cognizant that the differences in size and Lewis acidity of these alkali metal ions, however, may affect other phase transition states and are contingent on the concentration (or technically as 'molar ratio'). Such details are clearly beyond the scope of the present work and will be subjects of future work. To evaluate the thermal stability of KTFSA-based ionic liquid, thermal gravimetric and differential thermal analyses (TG-DTA) were performed. TG-DTA curves for 0.5 M KTFSA/$Pyr_{13}$TFSA under a nitrogen atmosphere are shown in **Figure S2** with that of 0.5 M KFSA/$Pyr_{13}$FSA which has recently been reported as a feasible electrolyte candidate for potassium-ion (K-ion) battery.[19] Thermal decomposition temperatures (upon a 5% loss in weight) are 690 K (417 °C) for 0.5 M KTFSA/$Pyr_{13}$TFSA and 580 K (307 °C) for 0.5 M KFSA/$Pyr_{13}$FSA, respectively, indicating that 0.5 M KTFSA/$Pyr_{13}$TFSA ionic liquid is more advantageous in the high-temperature operation of K-ion batteries.

Other pertinent macroscopic features of KTFSA-based ionic liquid, such as densities, viscosities and ionic conductivities dependence on temperature, were measured at various concentrations of KTFSA. **Figures 2a and 2d** illustrate the temperature dependence on densities of KTFSA/$Pyr_{13}$TFSA and KTFSA/EMITFSA ionic liquid systems. The density ($\rho$) of each ionic liquid varied linearly with the temperature and was nicely fitted using the following equation:

$$\rho = a + bT \tag{1}$$

where $T$, $b$ and $a$, denote the temperature (K), the coefficient of volume expansion (g cm$^{-3}$ K$^{-1}$) and the density at 0 K (g cm$^{-3}$), respectively. The fitted parameters are listed in **Table S1**.

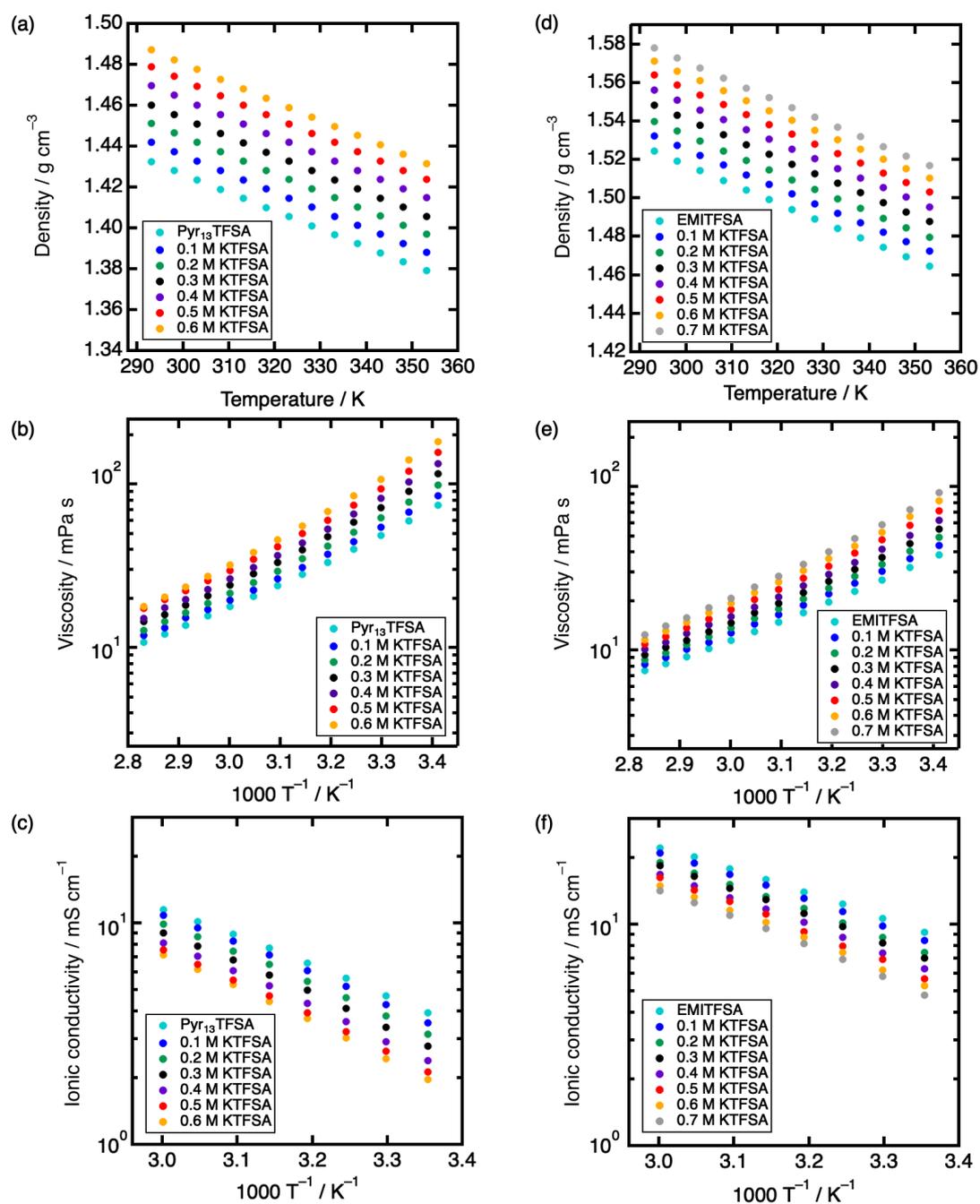

**Figure 2.** Physicochemical properties of potassium bis(trifluoromethanesulfonyl)amide (KTFSA)-based ionic liquids. Plots showing the temperature dependence on density (**(a)** and **(d)**), viscosity (**(b)** and **(e)**), and ionic conductivity (**(c)** and **(f)**) for KTFSA-based ionic liquids using 1-methyl-1-propylpyrrolidinium TFSA (Pyr$_{13}$TFSA) and 1-ethyl-3-methyl imidazolium TFSA (EMITFSA) with various KTFSA salt concentrations.

The density increased with the increase in concentration of KTFSA. Furthermore, the densities of KTFSA-based ionic liquid are compared to those of LiTFSA- and NaTFSA-based ionic liquids, as succinctly shown in **Figures S3a and S3d**. KTFSA-based ionic liquids exhibit the highest density, which is not unprecedented considering that the mass of potassium cation is greater than that of lithium and sodium. **Figures 2b and 2e** show the Arrhenius plots of the viscosities of KTFSA/Pyr$_{13}$TFSA and KTFSA/EMITFSA ionic liquid systems. The temperature dependence on the viscosity ($\eta$) is principally expressed by the Vogel-Tamman-Fulcher (VTF) equation as provided below: [23]

$$\eta = A_\eta exp\left(\frac{B_\eta}{T-T_0}\right) \qquad (2)$$

where $T_0$, $B_\eta$ and $A_\eta$ denote, respectively, the ideal glass transition temperature (K), the constant relating to Arrhenius activation energy for the viscous behavior (K) and the scaling factor. The fitted parameters are furnished in **Table S2**. Viscosity increases with increasing concentration of KTFSA, presumably due to the formation of relatively large ion pairs comprising potassium cations and TFSA anions. Generally, the ionic conductivity of ionic liquids can be correlated to their viscosity, as dictated by the Walden rule.[24] Arrhenius plots of the ionic conductivity of KTFSA/Pyr$_{13}$TFSA and KTFSA/EMITFSA ionic liquid systems are exhibited in **Figures 2c and 2f**. The temperature dependence on the ionic conductivity ($\sigma$) was also fitted by the following equation:

$$\sigma = A_\sigma exp\left(\frac{-B_\sigma}{T-T_0}\right) \qquad (3)$$

where $T_0$, $B_\sigma$ and $A_\sigma$ represent the ideal glass transition temperature (K), the constant pertaining to Arrhenius activation energy for the conduction behavior (K) and the scaling factor, respectively. The fitted parameters are provided in **Table S3**. Unlike the trend observed in the density, the ionic conductivity decreases upon increasing the concentration of KTFSA. It is worthy to mention that ionic liquids consisting of 0.5 M KTFSA/Pyr$_{13}$TFSA and 0.5 M KTFSA/EMITFSA exhibit conductivity of 2.1 mS cm$^{-1}$ and 5.7 mS cm$^{-1}$ at 298 K, respectively. Although we are aware that the attained ionic conductivities of these ionic liquids are lower than those of organic electrolytes, [25, 26] still they can withstand reliable battery applications.

To further evaluate the transport properties of KTFSA-based ionic liquids, a comparison was made with those of LiTFSA- and NaTFSA-based ionic liquids which possess a concentration of 0.5 M (as is shown in **Figures S3b**, **S3c**, **S3e**, **and S3f**). The fitted parameters, based on Vogel-Tamman-Fulcher (VTF) equation, and pertinent values at 298 K for each ionic liquid are furnished in **Tables S2**, **S3**, and **S4**. In the case of Pyr$_{13}$TFSA, 0.5 M KTFSA/Pyr$_{13}$TFSA displays lower viscosity and higher ionic conductivity. On the contrary, in the case of EMITFSA, 0.5 M KTFSA/EMITFSA exhibits slightly higher ionic conductivity, whereas the viscosity was almost the same. These transport properties are considered to be largely related to the solvation state of alkali metal ions in the ionic liquid. It is well known that TFSA$^-$ anions serve as bidentate ligands for metal ions using oxygen atom in sulfonyl group. Seminal studies, based on Raman spectroscopy and molecular dynamic (MD) simulations, have indeed revealed that the number of solvated TFSA$^-$ anions are 2 for Li$^+$, and 3 for both Na$^+$ and K$^+$.[27] In Pyr$_{13}$TFSA, however, there is a significant difference in the viscosity and ionic conductivity between 0.5 M KTFSA/Pyr$_{13}$TFSA and 0.5 M NaTFSA/Pyr$_{13}$TFSA, albeit the number of solvated TFSA$^-$ anions are the same. These results can be rationalized to arise from the difference in the ionic radii of Na (102 pm) and K (138 pm),[28] in addition to the Lewis acidity which determine the solvation structure. Nevertheless, it is irrefutable that the domain and periodic structure of ionic liquids also do impact. Worth mentioning is that Borodin and co-workers have reported recently that the binding energy, the size of first coordination shell and the exchange rate of TFSA$^-$ anion around metal cations in ionic liquid have a profound effect on the transport properties using MD simulations.[29] Although beyond the scope of the present study, further assessment of the intriguing transport properties of the KTFSA-based ionic liquids calls for rigorous experimental and theoretical approaches such as MD simulations and neutron / X-ray scattering. We await invigorating discussions from both theorists and experimentalists.

To investigate further on the potential of KTFSA-based ionic liquid as an electrolyte for battery applications, it was crucial to investigate its electrochemical properties. **Figure 3** displays the voltammograms of 0.5 M KTFSA/Pyr$_{13}$TFSA at room temperature along with those of 0.5 M LiTFSA/Pyr$_{13}$TFSA and 0.5 M NaTFSA/Pyr$_{13}$TFSA.[30] As apparent in **Figure 3**, a couple of redox peaks are observed at –3.3 V *versus* (vs.) ferrocene (Fc/Fc$^+$) in 0.5 M KTFSA/Pyr$_{13}$TFSA. These peaks are typical of the deposition and dissolution of potassium metal. The potential of deposition and dissolution of potassium metal was essentially the same as that in KFSA/Pyr$_{13}$FSA system. As for 0.5 M LiTFSA/Pyr$_{13}$TFSA and 0.5 M LiTFSA/Pyr$_{13}$TFSA, peaks corresponding to the deposition and dissolution of lithium and sodium metal are salient at –3.1 and –3.0 V vs. Fc/Fc$^+$ as explicitly shown in **Figure 3**. The voltage stability limits (technically, electrochemical windows) of each ionic liquid are summarized in **Table 1** along with their anodic stability using Pt as a working electrode. The coulombic efficiency of deposition and dissolution for 0.5 M KTFSA/Pyr$_{13}$TFSA was relatively low compared with that of 0.5 M LiTFSA/Pyr$_{13}$TFSA, suggesting that the adhesion of the deposited potassium metal on the Ni electrode is distinct from that of lithium metal. As for 0.5 M KTFSA/EMITFSA, *vide infra*, peaks attributable to the deposition of potassium metal are not observed on the electrode owing to the lower reduction stability of EMI$^+$ cations, as is shown in **Figure S4**. In view of the wide electrochemical window in conjunction with the feasibility of reversible deposition and dissolution of potassium metal, KTFSA/ Pyr$_{13}$TFSA ionic liquids are envisaged as promising electrolyte candidates for potassium-ion battery.

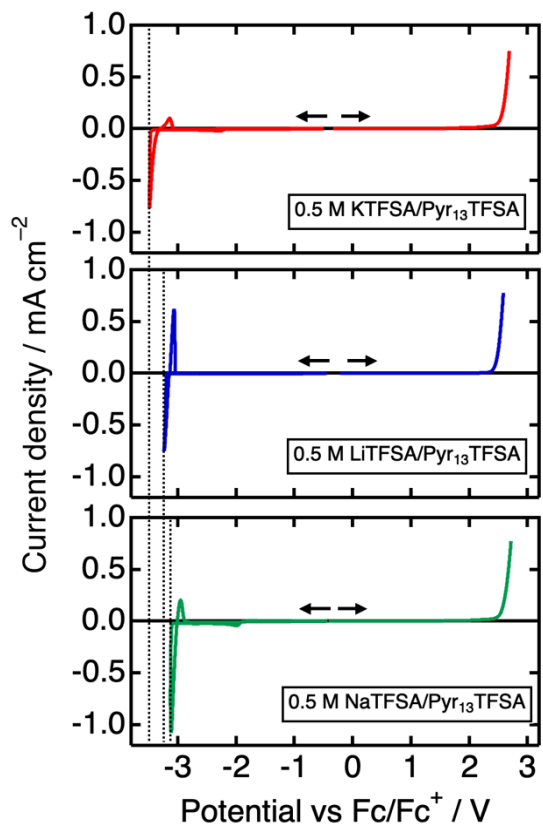

**Figure 3.** Electrochemical stability of TFSA-based ionic liquids containing alkali metal ion at 298 K. Cyclic and linear voltammograms of 0.5 M KTFSA/Pyr$_{13}$TFSA, 0.5 M LiTFSA/Pyr$_{13}$TFSA, and 0.5 M NaTFSA/Pyr$_{13}$TFSA at 298 K. The potential is shown based on ferrocene (Fc/Fc$^+$) as reference. Note that the working electrode was, respectively, Ni (cathodic limit) and Pt (anodic limit). Cyclic voltammetry (technically abbreviated as CV) and linear sweep voltammetry (LSV) were conducted to assess the cathodic limit and anodic limit, respectively. Scan rate was set at 1 mV s$^{-1}$. Dashed lines indicate the cathodic limit of each ionic liquid. 0.5 M KTFSA/Pyr$_{13}$TFSA ionic liquid exhibits a large electrochemical window compared with others, as detailed in **Table 1**.

**Table 1.** Voltage stability limits of bis(trifluoromethanesulfonyl) amide (TFSA)-based ionic liquids with potassium, sodium and lithium metal cations. The table summarizes the limit potentials and electrochemical windows of the ionic liquids. For the sake of clarity, the working electrode was, respectively, Ni (cathodic limit) and Pt (anodic limit). Scan rate was set at 1 mV s$^{-1}$.

| Ionic liquids | Limit potential vs. Fc/Fc$^+$ / V [a] | | Electrochemical window / V |
| --- | --- | --- | --- |
| | Cathode | Anode | |
| 0.5 M KTFSA / Pyr$_{13}$TFSA | −3.48 | 2.53 | 6.01 |
| 0.5 M LiTFSA / Pyr$_{13}$TFSA | −3.20 | 2.45 | 5.65 |
| 0.5 M NaTFSA / Pyr$_{13}$TFSA | −3.11 | 2.55 | 5.66 |

[a] The potential at which a current density of −0.1 mA cm$^{-2}$ was observed.

To validate more on the superiority of KTFSA/Pyr$_{13}$TFSA as electrolyte candidates for potassium-ion battery, the anode properties in the electrolyte was evaluated using potassium symmetric cells. **Figure 4** shows the voltage changes during galvanostatic deposition and dissolution of potassium using potassium symmetric cells consisting of 0.5 M KTFSA/Pyr$_{13}$TFSA and 0.5 M KTFSA/EMITFSA. Although the voltage changes asymmetrically and surpasses 1.0 V after 20 hours in 0.5 M KTFSA/EMITFSA, the deposition and dissolution of potassium is relatively stable with low overvoltage in 0.5 M KTFSA/Pyr$_{13}$TFSA ionic liquid. These results bring to the fore ionic liquids based on KTFSA/Pyr$_{13}$TFSA as propitious for potassium-ion battery application, particularly when coupled to high-voltage cathode materials.

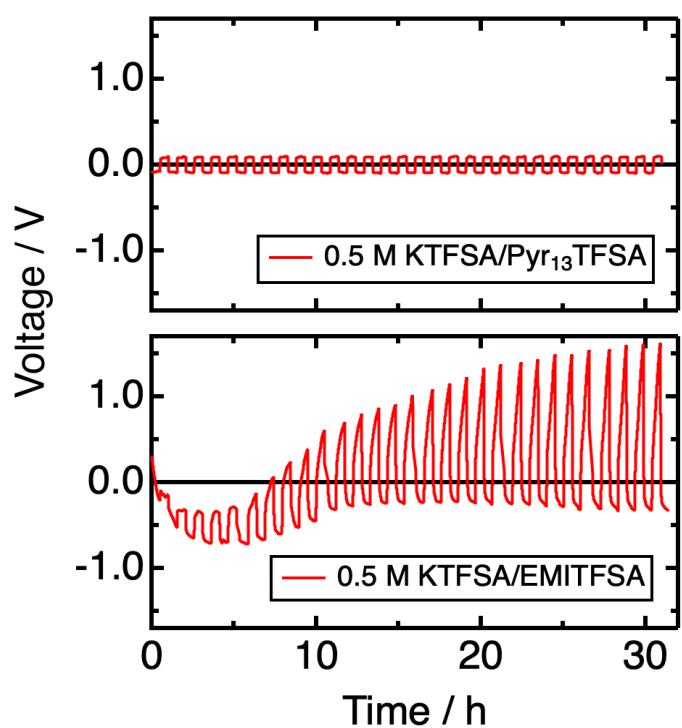

**Figure 4.** Evaluation of voltage polarization in potassium symmetric cells using KTFSA-based ionic liquid. Voltage profiles during galvanostatic deposition and dissolution of potassium in potassium symmetric cells using 0.5 M KTFSA/Pyr$_{13}$TFSA and 0.5 M KTFSA/EMITFSA at 298 K. Current applied for deposition and dissolution was 5 μA for 0.5

h (Current density was calculated to be equivalent to 6.4 µA cm$^{-2}$). It is discernible that the deposition and dissolution of potassium in 0.5 M KTFSA/Pyr$_{13}$TFSA was quite stable with relatively low overvoltage.

## 2.2. Synthesis of New High-Voltage Layered Compounds: K$_2$Ni$_{2-x}$Co$_x$TeO$_6$ ($x$ = 0.25, 0.5 and 0.75)

The rationale behind the syntheses of cobalt (Co)-substituted (Co-doped) nickel orthotellurate layered compositions, is based on our preliminary studies that show that partial Co-substitution in K$_2$Ni$_{2-x}$Co$_x$TeO$_6$ leads to a profound increase in the attained average voltage.[8] New orthotellurate K$_2$Ni$_{2-x}$Co$_x$TeO$_6$ (where $x$ = 0.25, 0.5 and 0.75) layered compounds were synthesized by the conventional high-temperature solid-state ceramics route as detailed in the **Experimental** section. **Figure 5** shows the refined X-ray diffraction (XRD) patterns of the new orthotellurate compounds, with the calculated and observed diffraction patterns indicated in black and red, respectively, whereas the differences are plotted in blue. All the diffraction peaks were satisfactorily indexable to a centrosymmetric hexagonal space group *P*6$_3$/*mcm*, as further evinced by high-resolution transmission electron microscopy (TEM), and more specifically, selected area electron diffractograms (see inset of **Figures 5a**, **5b and 5c,** and Supplementary Information **(Figures S5, S7 and S9)**). The refinement results (*i.e.*, the deduced atomic coordinates and lattice parameters are shown in the **Supplementary Information** section (**Tables S6, S7 and S8**). K$_2$Ni$_{2-x}$Co$_x$TeO$_6$ (where $x$ = 0.25, 0.5 and 0.75) adopt layered frameworks as depicted in inset of Figure 1a. The present orthotellurate layered system comprises TeO$_6$ and Ni/CoO$_6$ octahedra in a honeycomb fashion, spanning over the *ab*-plane, and mobile K$^+$ ions located in a plane between the octahedral layers. Morphological aspects of the as-prepared compounds through scanning electron microscopy (SEM), further reveals homogeneous particle distribution with a mean size in the range of 1~3 µm (shown in **Figure S6**, **S8**, **and S10** in the **Supplementary Information**). Moreover, the corresponding energy-dispersive X-ray (EDX) spectroscopic mappings show clearly overlapping signals of K, Te, Ni and Co in $K_2Ni_{2-x}Co_xTeO_6$, which demonstrate an even incorporation of Co into the compounds and further reveal fine elemental homogeneities. The composition of the samples was further validated by inductively-coupled plasma (ICP) measurements, as adumbrated in **Table S5**. Thermal analyses show that the new orthotellurate layered compositions are stable up to 1073 K (800 °C), which indicates good thermal stability that is pivotal for high-performance cathode materials (see **Figure S11**).

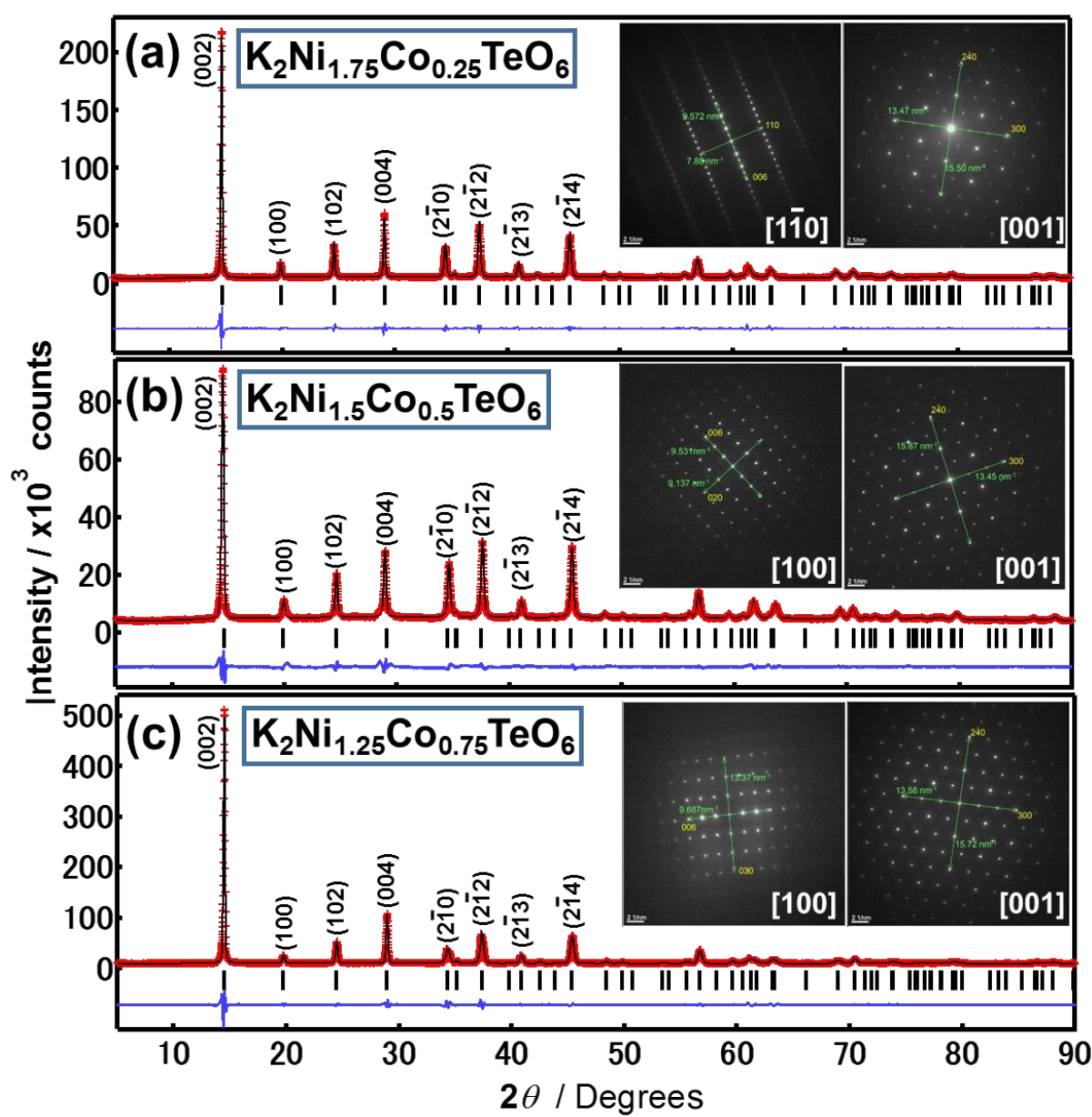

**Figure 5.** Synthesis and characterization of high-voltage orthotellurate compositions. Rietveld refinement of the powder X-ray diffraction (XRD) patterns of $K_2Ni_{2-x}Co_xTeO_6$ ($x$ = 0.25, 0.5, 0.75) for which high resolution data was available. The wavelength was maintained at Cu-$K\alpha$. Details regarding the refinement are furnished in the **Experimental** and **Supplementary**

**Information** section. The observed and calculated peaks are shown in red and black, respectively. The difference between the calculated and observed intensity is indicated in blue, while black ticks indicate the position of the Bragg peaks of the orthotellurate phase indexed in the $P6_3/mcm$ hexagonal space group as is shown in the transmission electron microscopy (TEM) inset images. For clarity, more enlarged electron diffraction images are appended in the **Supplementary Information** (**Figures S5, S7 and S9**).

## 2.3. Electrochemistry of New High-Voltage Layered Compounds Coupled with KTFSA-Based Ionic Liquids

High voltages can be envisaged with the utilization of the new orthotellurate $K_2Ni_{2-x}Co_xTeO_6$ layered compositions. Moreover, the presence of $K^+$ cations sandwiched between honeycomb metal slabs, can often have intriguing electrochemical behavior. Therefore, we set out to assess their electrochemistry with TFSA-based ionic liquids that display a wide voltage window. **Figure 6** shows the electrochemical responses of $K_2Ni_{2-x}Co_xTeO_6$ ($x$ = 0.25, 0.5, 0.75) as studied by means of cyclic voltammetry in the voltage ranges between 2.7 V and 4.5 V. Voltammograms for other tellurate compositions are furnished in the **Supplementary Information** section (**Figure S12**). In all compositions tested, main reduction and oxidation (redox) waves (peaks) around 4 V are evident. The redox waves neatly superimpose on subsequent cycling (not shown) indicating excellent reversibility of the high-voltage cathodes in ionic liquids based on KTFSA. Note that the overvoltage emanating from the reaction on the potassium metal as a negative electrode is sufficiently low, as apparent in **Figure 4**. In addition, the peaks are quite broad indicative of a single-phase reaction upon $K^+$ extraction / insertion in the layered $K_2Ni_{2-x}Co_xTeO_6$ orthotellurate compositions.

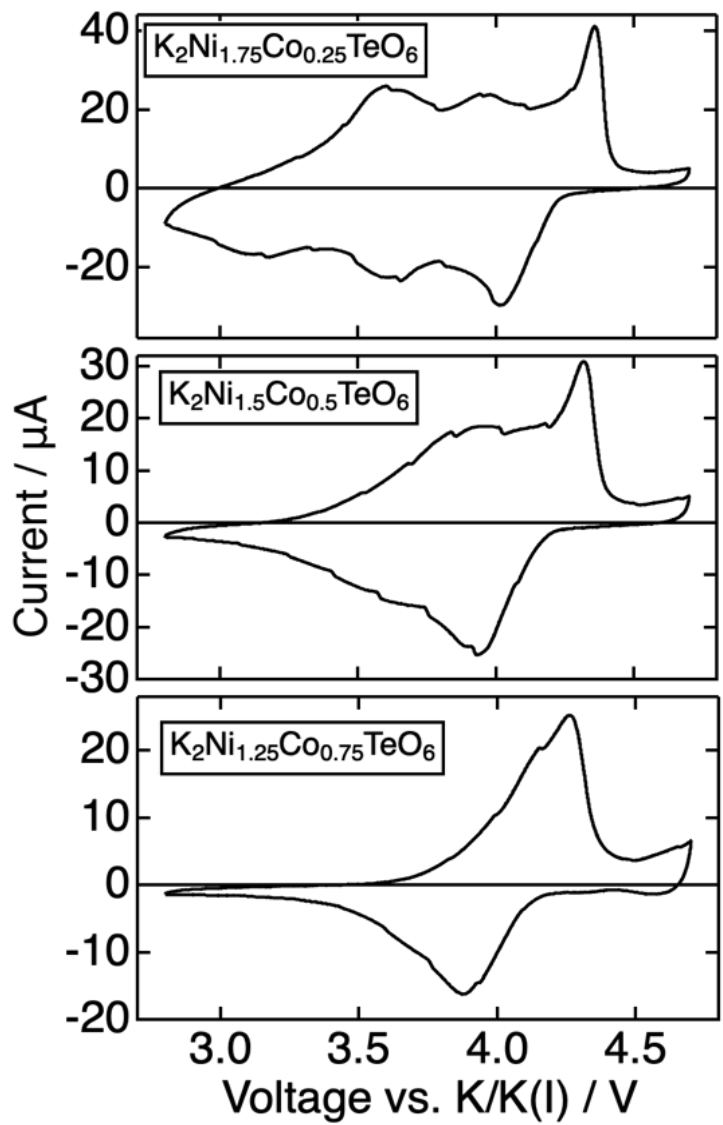

**Figure 6.** Voltage responses of new high-voltage orthotellurate layered compositions. Cyclic voltammograms of $K_2Ni_{2-x}Co_xTeO_6$ ($x$ = 0.25, 0.5, 0.75) in K half-cells using 0.5 M KTFSA/Pyr$_{13}$TFSA ionic liquid, showing the main redox voltage peaks centered around 4 V. Measurements were done at 298 K under a scan rate of 0.1 mV s$^{-1}$. Cyclic voltammograms for other tellurate compositions are provided in **Figure S12**.

More details on the cathode electrochemistry of some of these honeycomb-layered $K_2Ni_{2-x}Co_xTeO_6$ orthotellurates was further probed by electrochemical cycling obtained in a galvanostatic mode. **Figures 7a, 7b and 7c** show the electrochemical behavior of $K_2Ni_{2-x}Co_xTeO_6$ ($x$ = 0.25, 0.5 and 0.75) as explored in potassium half-cells using 0.5 M KTFSA in Pyr$_{13}$TFSA ionic liquid as an electrolyte. The voltage-composition traces of $K_2Ni_{2-x}Co_xTeO_6$ ($x$ = 0.25, 0.5 and 0.75) indicate reversible K$^+$ extraction and insertion at mean voltages of 3.7 V, 3.85 V and 4 V, respectively. In addition, differential capacity-voltage plots are shown in **Figure S13**, indicating good agreement with the results of the cyclic voltammograms displayed in **Figure 6**. Voltage-capacity plots and the corresponding performance upon long-term cycling (100 cycles) are also summarized in the **Supplementary Information** section (**Figure S14**). The reversible capacities corresponding to removal of ~0.48 K$^+$ ions (solely based on electron counting), are sustained upon subsequent cycling (in the case of K$_2$Ni$_{1.75}$Co$_{0.25}$TeO$_6$). Reversible capacities are attained upon cycling at high-voltage regimes that are compatible with ionic liquids, indicative of good structural integrity of the layered cathode frameworks.

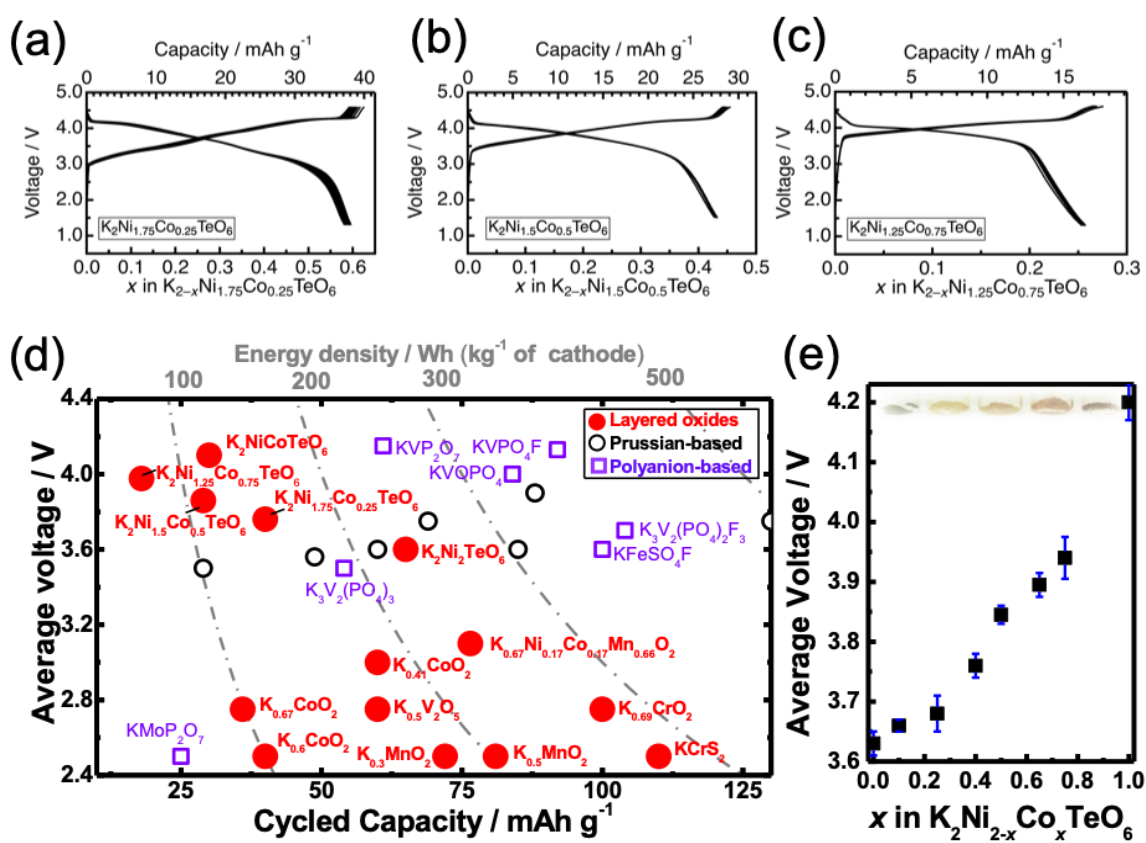

**Figure 7.** Electrochemical performance of new high-voltage orthotellurate layered compositions K$_2$Ni$_{2-x}$Co$_x$TeO$_6$ ($x$ = 0.25, 0.5, 0.75). Voltage-capacity plots of **(a)** K$_2$Ni$_{1.75}$Co$_{0.25}$TeO$_6$, **(b)** K$_2$Ni$_{1.5}$Co$_{0.5}$TeO$_6$ and **(c)** K$_2$Ni$_{1.25}$Co$_{0.75}$TeO$_6$ at current density corresponding to C/20 (Here 1 C corresponds to 127.86 mA g$^{-1}$, 127.84 mA g$^{-1}$ and 127.82 mA g$^{-1}$, respectively) in K half-cells using 0.5 M KTFSA/Pyr$_{13}$TFSA ionic liquid demonstrating stable performance at high voltages. The 2nd to the 10th charge and discharge curves are shown, for brevity. **(d)** Ragone plot benchmarking the mean (average) voltages of K$_2$Ni$_{2-x}$Co$_x$TeO$_6$ ($x$ = 0.25, 0.5, 0.75) with some of the reported cathode materials. **(e)** Trend in the variation of the mean voltages attained in K$_2$Ni$_{2-x}$Co$_x$TeO$_6$ ($x$ = 0, 0.1, 0.25, 0.5, 0.75, 1) layered orthotellurate

compositions, showing the feasibility of further tuning the voltage responses through compositional changes in the layered tellurates. Inset shows the various colors of selected powders.

**Figure 7d** shows a comparative plot of the performance of $K_2Ni_{2-x}Co_xTeO_6$ ($x$ = 0.25, 0.5 and 0.75) new compositions along with the reported cathode materials for rechargeable potassium-ion battery. Although the attained capacity is relatively low, the layered orthotellurate $K_2Ni_{2-x}Co_xTeO_6$ compositions benchmark the highest voltages for layered cathodes so far reported. High voltages can still be anticipated with further substitution of cobalt, as envisaged in **Figure 7e**, or with partial substitution of other congener transition metal ions. Ample room exists for improvement of capacity at elevated temperatures, which is an advantage when the high-voltage cathodes are coupled with TFSA-based ionic liquids that are stable at elevated temperatures.

## 2.4. Structural Stability of New High-Voltage Layered Compounds Coupled with KTFSA-Based Ionic Liquids upon Cycling

The structural stability upon electrochemical charging and discharging (cycling) was further investigated by *ex situ* X-ray diffraction (XRD) analyses of the (dis)charged layered $K_2Ni_{2-x}Co_xTeO_6$ electrodes. We focused on $K_2Ni_{1.75}Co_{0.25}TeO_6$ composition as it presents the highest reversible capacity amongst the orthotellurate compositions tested. Structural evolution for $K_2Ni_{1.5}Co_{0.5}TeO_6$, *ex situ* XRD data of which we also measured, are furnished in the **Supplementary Information** section (**Figure S15**). The evolution of the XRD pattern obtained by *ex situ* XRD study during the initial (dis)charge cycle is shown in **Figure 8a**. At first glance, no conspicuous changes can be seen from the XRD patterns upon charge and discharge, indicating that the layered framework is retained upon $K^+$ extraction and insertion. $K^+$ extraction from the interslab spaces of layered frameworks, in principle, leads to expansion along the axis perpendicular to the slabs (in this case, the *c*-axis) owing to increased repulsion between the slabs containing transition metals. **Figure 8b** shows an enlarged XRD pattern indicating shifts of Bragg diffraction peaks, and more specifically, the (00*l*) peaks that are very sensitive to $K^+$ (de)insertion. In particular, the (004) Bragg peaks generally shift to low and high angles upon charge and discharge, respectively, revealing that the *c*-axis expands (and contracts) with $K^+$ extraction (and insertion). **Figure 8c** depicts the overall reversible trend in the *c*-lattice parameter evolution, indicating that a highly repeatable topotactic $K^+$ extraction and insertion occurs in these high-voltage layered orthotellurate cathodes — an epitome being $K_2Ni_{1.75}Co_{0.25}TeO_6$ and $K_2Ni_{1.25}Co_{0.75}TeO_6$. We also are yet to ascertain the possible transition

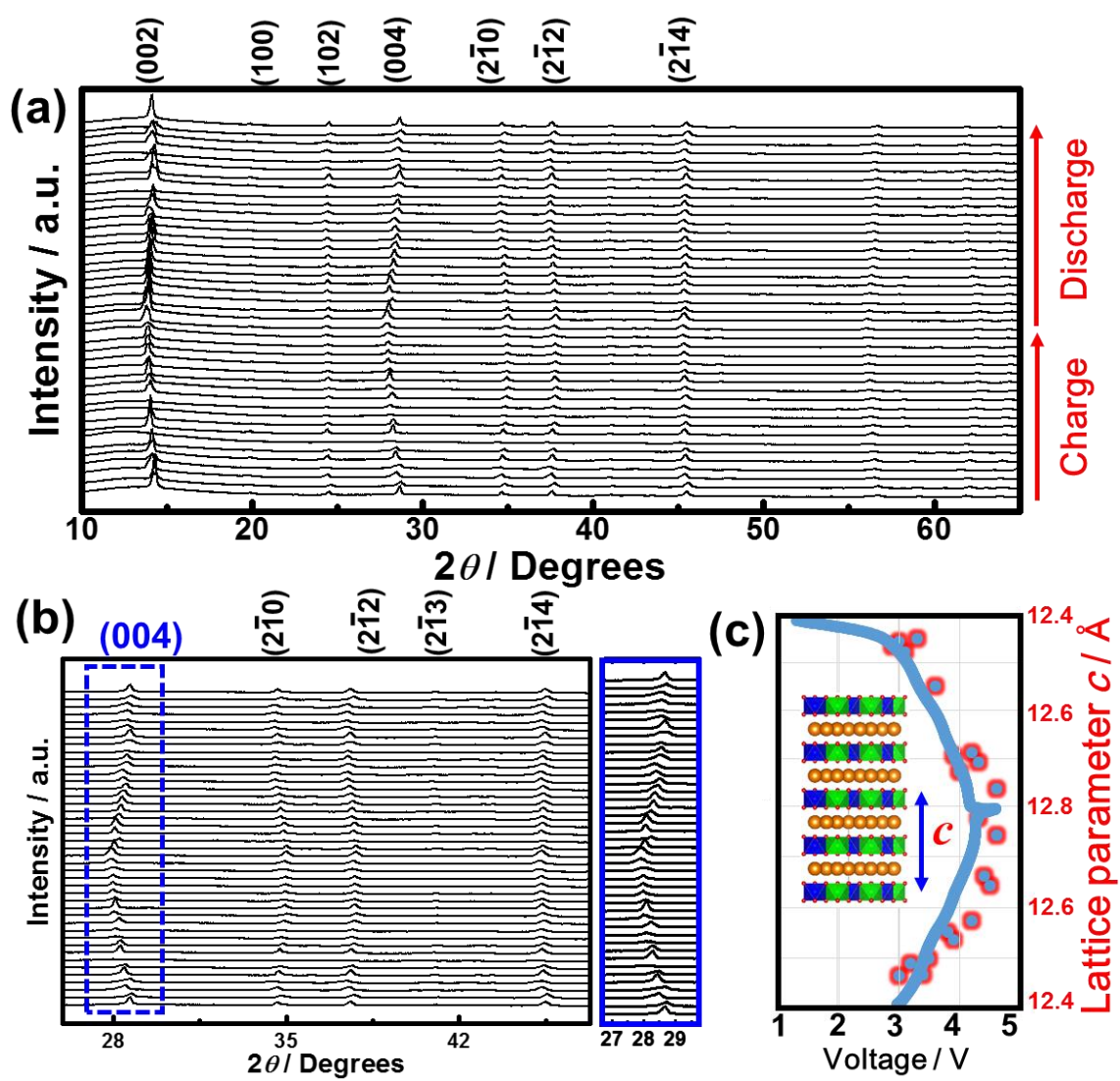

**Figure 8.** Crystal structural evolution of $K_2Ni_{2-x}Co_xTeO_6$ (*x* = 0.25) representative solid solution phase. **(a)** XRD *ex situ* patterns while $K_2Ni_{1.75}Co_{0.25}TeO_6$ is (dis)charged at a current

density commensurate to C/20 rate (20 h of (dis)charge; 1 C = 127.86 mA g$^{-1}$), **(b)** enlarged XRD pattern indicating peak shifts of the more sensitive (004) Bragg diffraction peaks during (dis)charging and **(c)** trend in variation of the lattice constants along the *c*-axis of the honeycomb layered framework upon charge (K$^+$ extraction) and discharge (K$^+$ re-insertion). The wavelength was set at Cu-*Kα*.

metal ions that take part in the redox process during K$^+$ extraction and insertion in these new orthotellurate compositions. A systematic study using synchrotron X-ray measurements is demanded, as we previously pursued with K$_2$Ni$_2$TeO$_6$.[25] Such details are clearly beyond the scope of the current study.

To recapitulate, this work has implications in the advancement of high energy density potassium-ion battery, particularly for electrolytes to match high-voltage cathode materials. Moreover, the prevailing wisdom with this nascent technology is that the sluggish potassium-ion kinetics tend to hinder the full utilization of apparently 'electrochemically active' cathode materials. It is hoped that this work will turn out to be trendsetting in the development of high-performance potassium-ion batteries also operable at intermediate temperatures using cathodes (whether quixotic or pragmatic) that were apparently perceived to be 'electrochemically inactive' at room temperature.

## 3. Conclusion

This study has elucidated ionic liquids based on potassium bis(trifluoromethanesulfonyl)amide (KTFSA) in 1-methyl-1- propylpyrrolidinium bis(trifluoromethanesulfonyl)amide (Pyr$_{13}$TFSA) as suitable and safe electrolytes for high-voltage cathode materials. Electrochemical measurements reveal 0.5 M KTFSA in Pyr$_{13}$TFSA ionic liquid electrolyte to exhibit a lower redox potential than both lithium and sodium. This finding reconciles the emerging knowledge that indeed a high-voltage battery configuration is feasible with potassium-ion technology. The suitability of these KTFSA-based ionic liquids has been shown by matching with new high-voltage layered cathode materials (*viz.*, $K_2Ni_{2-x}Co_xTeO_6$ or equivalently as $K_{2/3}Ni_{(2-x)/3}Co_{x/3}Te_{1/3}O_2$ ($x = 0.25, 0.5$ and $0.75$)), demonstrating good stability at high voltages. Further optimization of the performance of the layered cathodes remain; for instance, assessment of their performance at high temperatures within the range of practical interests. Nevertheless, the wide electrochemical window (around 6.0 V (within the stipulated conditions)) surmises that a 5 V-class potassium-ion battery configuration may not be far from reality. It is prudent to mention here that ionic liquids generally tend to be expensive, an aspect that may circumvent their widespread commercial deployment in battery applications. However, with bespoke manufacturing protocols and greater economies of scale (for instance, as governed by well-established market cost parameters such as the Herfindahl-Hirschman index), ionic liquids can be prepared on pilot scale at affordable costs.

A pending issue relates to the relatively high viscosity (and low ionic conductivity) exhibited by the ionic liquids at room temperature (*ca.* 120 mPa s at 25 °C (298 K) for 0.5 M KTFSA in $Pyr_{13}TFSA$), which may compromise battery performance metrics such as power density. Manipulating the ionic salt chemistry may be a promising route. For example, we have empirically shown that KTFSA incorporating imidazolium-based ionic liquid can afford low viscosities and subsequently high ionic conductivity, albeit at an expense of lowering the reductive stability of the electrolyte. There is ample room for further advancement in the performance of these ionic liquids, a research pursuit that demands unremitting efforts from the wide-spanning scientific community ㅡ electrochemists, theorists, material chemists and physicists. It is hoped that this work will spur extensive research on ionic liquid salts (as 'iconic chemistries') for high-energy density potassium-ion energy storage systems.

## 4. Experimental Section

*Preparation of Electrolyte*: 1-methyl-1-propylpyrrolidinium bis(trifluoromethanesulfonyl)amide (abbreviated as $Pyr_{13}TFSA$ ($Pyr_{13}TFSI$ or $Pyr_{13}Tf_2N$)) (Kanto Chemicals (Japan), purity of 99.9%), 1-methyl-1-propylpyrrolidinium bis(fluoromethanesulfonyl)amide ($Pyr_{13}FSA$) (Kanto Chemicals (Japan), purity of 99.9%), 1-ethyl-3methylimidazolium bis(trifluoromethanesulfonyl)amide (EMITFSA) (iolitec, purity of > 99.5%), lithium bis(trifluoromethanesulfonyl)amide (LiTFSA) (Morita Chemical Industries, purity of > 99.0%), potasssium bis(trifluoromethanesulfonyl)amide (KTFSA) (Morita Chemical Industries, purity of > 99.0%), lithium bis(fluoromethanesulfonyl)amide (LiFSA) (Tokyo Chemical Industries (TCI (Japan))), purity of 98.0%) were dried *in vacuo* at 358 K or 373 K for 24 hours prior to use. Sodium bis(trifluoromethanesulfonyl)amide (NaTFSA) was synthesized by reacting, in principle, HTFSA with $Na_2CO_3$. Preparation of electrolytes was conducted in an argon-filled glove box equipped with a gas purification system (Miwa MFG Co. Ltd., MDB-1NKPS or MDB-2LKSTS-NW) whose oxygen and water ($H_2O$) contents were maintained to concentrations of below 1 part per million (ppm).

*Measurements of Physicochemical Properties of Electrolyte*: A series of viscosity measurements were carried out using a viscometer (EMS-1000S (Kyoto Electronics)), whilse densities were measured with a vibrating-type densitometer (DMA$^{TM}$ 4500 (Anton Paar)). Ionic conductivity was measured with an airtight four-probe conductivity cell, consisting of two inner platinum wire-electrodes (diameter of 1.0 mm) for monitoring the potential difference and two outer platinum disk-electrodes (with a diameter of 13.0 mm) for feeding an alternating current amplitude of 10 μA root-mean-square with a potentiostat machine (PARSTAT 2263 (Princeton Applied Research)) after the cell calibration using 0.1 M (mol dm$^{-3}$) KCl aqueous solution. Thermogravimetry-differential thermal analysis (TG-DTA) was conducted using a TG-DTA instrument (TG-DTA200 (Hitachi High-Tech.)) with a platinum ampoule in the temperature ranges of 303–1200 K at a ramp rate of 10 K min$^{-1}$ under flowing nitrogen (N$_2$). Differential scanning calorimetry (DSC) was performed using a calorimeter (DSC6220 (SII nanotechnology)) with a sealed-aluminum ampoule at a ramp rate of 5 K min$^{-1}$. Note that the samples for the aforementioned measurements were prepared in an argon-filled glove box. Electrochemical measurements were performed using customised three-electrode cells in the argon-filled glove boxes with a galvanostatic machine (HZ-Pro (Hokuto denko)). A nickel (2.01 × 10$^{-2}$ cm$^2$) or platinum (1.77 × 10$^{-2}$ cm$^2$) plate was used as a working electrode, which were polished using 0.05 μm alumina suspension prior to measurements. A platinum wire was used as a counter electrode. Additionally, a silver wire immersed in Pyr$_{13}$TFSA containing 0.1 M silver trifluoromethanesulphonate (AgCF$_3$SO$_3$) was used as a reference electrode. Upon completion of measurements, appropriate amounts of ferrocene was added to the electrolyte and cyclic voltammetry tests were duly conducted using platinum electrode in order to convert (or technically, calibrate) the potential from Ag/Ag(I) to ferrocene/ferrocenium (Fc/Fc$^+$).

*Material Synthesis*: Conventional solid-state reaction protocols were adopted to prepare new high-voltage layered orthotellurates with nominal compositions K$_2$Ni$_{2-x}$Co$_x$TeO$_6$ ($x$ = 0.25, 0.5 and 0.75). TeO$_2$ (Aldrich, purity of ≥99.0%), NiO (Kojundo Chemical Laboratory (Japan), purity of 99%) and K$_2$CO$_3$ (Rare Metallic (Japan), purity of 99.9%) were finely ground (using a pestle and an agate mortar) in stoichiometric proportions to yield K$_2$Ni$_{2-x}$Co$_x$TeO$_6$ precursors. The precursor powders were then pelletized and annealed in gold crucibles in air for 20−24 h (with intermediate grinding) in the temperature ranges of 750 °C (1023 K) to 800 °C (1073 K). Annealing was done at a ramp rate of +400 K h$^{-1}$ (or +400 °C h$^{-1}$) and then cooled to room temperature upon completion of the annealing dwell time. To avert any moisture exposure, the obtained powders were transferred to a glove box (argon-purged).

*X-ray Diffraction Analyses*: X-ray diffraction (XRD) measurements were performed using a diffractometer (Bruker D8 ADVANCE) employing Cu-$K\alpha$ radiation (*viz.*, $\lambda =1.54056$ Å). XRD protocols (in Bragg-Brentano geometry) were performed in a $2\theta$ range of 5° ~ 90° at a step size of 0.01°. XRD refinement and data analysis was carried out by the Rietveld procedure implemented in the JANA 2006 program, and the visualization of the crystal structure was done using VESTA crystallographic software.[31, 32] The structural refinements for $K_2Ni_{2-x}Co_xTeO_6$ ($x = 0.25$, 0.5 and 0.75) were performed on XRD data based on the $K_2Ni_2TeO_6$ structural model. Chebyshev polynomials to describe the background were used during the structural refinement, while pseudo-Voigt profile function was applied to correct peak asymmetry. Preferred orientation with respect to the (00*l*) axis was also considered for a final satisfactory fit.

XRD *ex situ* measurements (Cu-$K\alpha$ radiation) for $K_2Ni_{2-x}Co_xTeO_6$ ($x = 0.25$ and 0.5) were collected also in Bragg–Brentano geometry for electrodes cycled at different (dis) charge depths at a current density equivalent to C/20 rate. Le Bail profile fitting of the patterns was done to obtain the trend in lattice parameters evolution during (dis)charge at almost every 0.1 V voltage change.

*Morphological and Elemental Chracterization*: Scanning electron microscope (JSM-6510LA) was used to analyze the morphological aspects of the as-synthesized powders where mapping of the constituent elements was carried out using the energy dispersive X-ray (EDX) imaging function. TEM (Transmission Electron Microscopy) and high-resolution images using TEM (HRTEM) were obtained on a TITAN80-300F at an acceleration voltage of 200 kV without exposure to neither moisture nor air. HRTEM image simulations were conducted using the JEMS 31(PECD) software.

*Thermal Analyses*: Thermogravimetric and differential thermal analysis (TG-DTA) was done using a TG-DTA instrument 2020SA (Bruker AXS) in the temperature ranges of 303–1173 K at a ramp rate of 5 K min$^{-1}$ under flowing N$_2$ or Ar. As is customary, a baseline correction of the TG curve was carried out by measuring the empty Pt crucible prior to each measurement.

*Electrochemical Measurements*: All electrochemical measurements detailed herein were performed at room temperature. Coin cells assembly and related protocols were performed inside an Ar-purged glove box (MIWA, MDB-1KP-0 type) with oxygen and water (H$_2$O) contents below 1 ppm. Regarding to fabrication of the electrode, K$_2$Ni$_{2-x}$Co$_x$TeO$_6$ ($x$ = 0.25, 0.5 and 0.75) new cathodes and related solid-solution derivatives were intimately mixed with carbon and polyvinylidene fluoride (PVdF) binder to attain a weight formulation of cathode material : carbon : binder as 82 : 17 : 3. Mindful of the rheological properties that affect battery performance, a viscous slurry was made by suspending the mixture in *N*-methyl-2-pyrrolidinone (NMP), which was then cast on aluminum foil with a mass loading of 3–5 mg cm$^{-2}$. Composite cathodes were punched out and dried at 393 K (120 °C) *in vacuo*. Electrochemical performances were assessed in CR2032-type coin cells using K$_2$Ni$_{2-x}$Co$_x$TeO$_6$ ($x$ = 0.25, 0.5 and 0.75) composite cathode (and related solid-solution derivatives), separated from potassium metal anode (technically referred to as counter electrode) by glass fiber discs soaked with electrolyte. A 0.5 mol dm$^{-3}$ potassium bis(trifluoromethanesulfonyl)amide (abbreviated as KTFSA (KTFSI or KTf$_2$N depending on literature)) in 1-methyl-1-propylpyrrolidinium bis(trifluoromethanesulfonyl) amide (hereafter Pyr$_{13}$TFSA) (Kanto Chemicals (Japan), purity of 99.9%, <20 ppm H$_2$O) ionic liquid was used as the electrolyte. Galvanostatic cycling was done at a current rate corresponding to C/20 (20 being the hours required to (de)insert 2 K$^+$ per formula unit; 1C = ~128 mA g$^{-1}$). Unless otherwise stated, cyclic voltammetry tests were conducted between 2.8 V and 4.6 V (*vs*. K$^+$/K) and at a scan rate of 0.1 mV s$^{-1}$.

## Supporting Information

Supporting Information is available via the following link: https://chemistry-europe.onlinelibrary.wiley.com/action/downloadSupplement?doi=10.1002%2Fcelc.201900689&file=celc201900689-sup-0001-misc_information.pdf

## Acknowledgements

We gratefully acknowledge Dr. Hajime Matsumoto for his useful inputs in our discussions. We also acknowledge Ms. Kumi Shiokawa and Ms. Yumi Haiduka for their advice and technical help as we conducted the electrochemical and XRD measurements. This work was conducted in part under the auspices of the Japan Society for the Promotion of Science (JSPS KAKENHI Grant Numbers 19K15686 and 19K15685) and Japan Prize Foundation.